\documentclass[twocolumn,superscriptaddress,amsfont,amssymb,amsmath, showpacs,balancelastpage, nofootinbib]{revtex4-1}
\usepackage{graphicx,longtable,mathrsfs,color,array}
\usepackage[unicode=true,pdfusetitle,
bookmarks=true,bookmarksnumbered=true,bookmarksopen=true,bookmarksopenlevel=1,
breaklinks=false,pdfborder={0 0 0},backref=false,colorlinks=true]{hyperref}
\hypersetup{citecolor=blue,filecolor=blue,linkcolor=blue,urlcolor=blue}
\usepackage[usenames,dvipsnames]{xcolor}
\usepackage{amssymb,amsmath,mathtools,mathrsfs,enumitem}
\usepackage{epsfig,subfigure,placeins,float}
\usepackage{booktabs,longtable,ctable,multirow}
\usepackage{exscale,relsize}
\usepackage[normalem]{ulem}
\usepackage[T1]{fontenc}
\usepackage[utf8]{inputenc}
\usepackage{enumerate}
\usepackage{aas_macros}
\usepackage{times, mathptmx}
\newcommand{\be}{\begin{equation}}
\newcommand{\ee}{\end{equation}}

\newcommand{\MPl}{M_{\rm Pl}}
\newcommand{\Oo}{{{\cal O}(1)}}
\newcommand{\comment}[1]{}
\newcommand{\nn}{\nonumber}

\newcolumntype{C}[1]{>{\centering\let\newline\\\arraybackslash\hspace{0pt}}m{#1}}

\begin{document}
\title{Radiative stability and observational constraints on dark energy and modified gravity}

 \author{J.~Noller}
  \affiliation{Institute for Theoretical Studies, ETH Z\"urich, Clausiusstrasse 47, 8092 Z\"urich, Switzerland}
  \affiliation{Institute for Particle Physics and Astrophysics, ETH Z\"urich, 8093 Z\"urich, Switzerland}
  \author{A.~Nicola}
  \affiliation{Institute for Particle Physics and Astrophysics, ETH Z\"urich, 8093 Z\"urich, Switzerland} 
  \affiliation{Department of Astrophysical Sciences, Princeton University, Princeton, NJ 08544, USA}

\begin{abstract}
We investigate the radiative stability of Horndeski scalar-tensor theories with luminally propagating gravitational waves (as extensively discussed in the wake of GW170817) and show that in general there is a tension between obtaining observable deviations from General Relativity (GR) in cosmology and the requirement of radiative stability. Using this as a constraint, we discuss the subsets of theories that are capable of evading this conclusion and yielding observable, radiatively stable departures from GR. 
We find several classes of theories that can do so, recovering known cases and identifying several additional radiatively stable cases.
Finally, we also extract the cosmological signatures of two particularly well-motivated radiatively stable classes of theories: shift-symmetric theories and theories with a conformal coupling between the scalar and gravity.
We find that cosmological parameter constraints on dark energy and modified gravity parameters for both of these two classes, which we explicitly compute using data from the Planck, SDSS/BOSS and 6dF surveys, are significantly tightened with respect to generic Horndeski theories.
\end{abstract}

  \date{\today}
  \maketitle
\noindent \textbf{Introduction:} 
In the recent past, great progress has been made in understanding how to precision-test our current leading theory of gravity, General Relativity (GR). Since GR is the single consistent theory of a massless spin-2 field (implicitly restricting ourselves to Lorentz-invariant theories), testing for deviations from GR becomes equivalent to testing for new (light) gravitational degrees of freedom. 
Scalar-tensor (ST) theories are a minimal deviation from GR in this sense, introducing a single additional degree of freedom and asking how it may affect gravitational interactions. 

Accordingly, Horndeski gravity \cite{Horndeski:1974wa,Deffayet:2011gz}, the most general Lorentz-invariant ST action that gives rise to second order equations of motion (and is consequently free of an Ostrogradski-ghost instability by default), has recently been the main workhorse of research into alternative theories of gravity. 
It is described by the following action
\begin{eqnarray}\label{Horndeski_action}
S_H=\int \mathrm{d}^4x \sqrt{-g}\left\{\sum_{i=2}^5{\cal L}_i[\phi,g_{\mu\nu}]\right\},
\end{eqnarray}
where the ${\cal L}_i$ are scalar-tensor Lagrangians (for a scalar $\phi$ and a massless tensor $g_{\mu\nu}$) given by
\begin{align}
{\cal L}_3&= -G_3 [\Phi], \quad\quad {\cal L}_4=  G_4 R+G_{4,X}\left([\Phi]^2-[\Phi^2]\right) , \nonumber \\
{\cal L}_5&= G_5 G_{\mu\nu}\Phi^{\mu\nu}
-\frac{1}{6}G_{5,X}\left( [\Phi]^3
-3[\Phi^2][\Phi] +2[\Phi^3]
\right)  \,. \label{Horndeski_lagrangians}
\end{align} 
The $G_i$ are functions of a scalar field $\phi$ and its derivative via $X \equiv -\tfrac{1}{2}\nabla_\mu \phi \nabla^\mu \phi$, where $\nabla_\mu$ is the covariant derivative for $g_{\mu\nu}$, and we also have ${\cal L}_2= G_2(\phi,X)$. We have used the shorthand $\Phi^{\mu\nu} \equiv \nabla^\mu \nabla^\nu \phi$ and square brackets denote traces, i.e. $[\Phi] = \Phi_\mu^\mu, [\Phi^2] = \Phi^\mu_\nu \Phi^\nu_\mu$ etc.
Finally, $G_{i,\phi}$ and $G_{i,X}$ denote the partial derivatives of the $G_i$, with respect to $\phi$ and $X$ respectively. Four free functions ($G_{2}, G_3, G_4, G_5$) therefore completely characterise this theory.

Recently, and motivated by the near simultaneous detections of GW170817 and GRB 170817A \cite{PhysRevLett.119.161101,2041-8205-848-2-L14,2041-8205-848-2-L15,2041-8205-848-2-L13,2041-8205-848-2-L12}, it was shown in \cite{Baker:2017hug,Creminelli:2017sry,Sakstein:2017xjx,Ezquiaga:2017ekz} that
imposing luminal propagation of gravitational waves (GWs), $c_{GW} = c$, significantly reduces this theory space in a cosmological context, namely by eliminating $G_5$ and $G_{4,X}$. 
The resulting, restricted Horndeski theory is described by
\begin{align} \label{Horndeski_simple}
S=\int \mathrm{d}^4x \sqrt{-g}\left\{ G_2(\phi,X) -G_3(\phi,X)[\Phi] + G_4(\phi) R\right\},
\end{align}
where now there are only three free functions left ($G_{2}, G_3, G_4$) and we highlight that $G_4$ is a function of $\phi$ only. For previous work on $c_{\rm GW} = c$ constraints see \cite{Amendola:2012ky,Amendola:2014wma,Linder:2014fna,Raveri:2014eea,
Saltas:2014dha,Lombriser:2015sxa,Lombriser:2016yzn,Jimenez:2015bwa,Bettoni:2016mij,Sawicki:2016klv}. 
Note that the derivation of \eqref{Horndeski_simple} implicitly assumes a scale/time/energy-independent speed of gravitational waves. 
Since GW170817 probes energy scales much larger than those of late-universe cosmology, in a modified gravity context it in principle tests the (unknown) UV completion of the cosmological theory. Indeed \cite{deRham:2018red} argue that generic Lorentz-invariant UV completions will bring a potentially subluminal cosmological speed of GWs back to luminal for the frequencies observed for GW170817. We refer to \cite{Creminelli:2017sry,deRham:2018red} for a discussion of the naturalness of such a scenario, but here (following \cite{Baker:2017hug,Creminelli:2017sry,Sakstein:2017xjx,Ezquiaga:2017ekz}) we will explicitly assume a luminally propagating, scale-independent speed of GWs -- an assumption that will be probed directly by LISA and pulsar timing arrays \cite{deRham:2018red}.

While excluding higher derivative interactions associated with ${\cal L}_{4,5}$, \eqref{Horndeski_simple} still includes a wide class of theories. Consequently the purpose of this paper is two-fold: Firstly, we argue that (in a sense we will make precise) large classes of models within \eqref{Horndeski_simple} cannot yield cosmological deviations from GR that are within reach of current/near-future observations {\it and} (radiatively) stable, but also highlight several classes of theories, where this is possible. In particular we demonstrate the radiative stability of some non-shift symmetric subsets of Horndeski theories for the first time. Secondly, we focus on two particularly well-motivated subsets of \eqref{Horndeski_simple}, which allow departures from GR with attractive stability properties, showing that such departures are significantly more restricted in these subsets than for generic theories within \eqref{Horndeski_simple}.
\\

\noindent \textbf{Linear Cosmology:} 
Cosmological deviations from GR are especially tightly constrained at the level of linear perturbations and we would therefore like to linearly perturb \eqref{Horndeski_simple} around an FRW background solution. The result is well-known for general Horndeski theories and, restricting to \eqref{Horndeski_simple}, the freedom in such a linearly perturbed action can be concisely parametrised in terms of four independent and free functions \cite{Bellini:2014fua} 
: The Hubble rate $H$ that controls the background expansion, the running of the Planck mass $\hat{\alpha}_M$, the kineticity $\hat{\alpha}_K$ (
essentially a proxy for the scalar speed of sound) and the braiding $\hat{\alpha}_B$ that quantifies kinetic mixing between metric and scalar perturbations. 
Note that the resulting linear action is equivalent to the one obtained in effective field theory (EFT)/effective action approaches for ST dark energy/modified gravity up to 2nd order in derivatives \cite{Gleyzes:2013ooa,Lagos:2016wyv,Lagos:2017hdr}. 

In this EFT spirit and motivated by the observed proximity of the background expansion to $\Lambda{\rm CDM}$, in what follows we will follow the minimal approach of \cite{BelliniParam,Alonso:2016suf} and fix the background to be that of $\Lambda{\rm CDM}$, considering and constraining perturbations around it. 
$\hat\alpha_K$, evaluated for \eqref{Horndeski_simple}, then satisfies
\begin{align}
H^2M^2\hat{\alpha}_K&=2X\left(G_{2,X}+2XG_{2,XX}-2G_{3,\phi}-2XG_{3,\phi X}\right) \nn \\
&+12\dot{\phi}XH\left(G_{3,X}+XG_{3,XX}\right),
\end{align}
where from \eqref{Horndeski_simple} the effective Planck mass $M^2$ can be read off to be $M^2=2 G_4$. However, $\hat \alpha_K$ is known to only very weakly affect cosmological observables \cite{BelliniParam,Alonso:2016suf}, which is linked to the fact that it drops out in the quasi-static approximation \cite{Bellini:2014fua}, so it can essentially be fixed to a 
fiducial value without affecting constraints. The remaining $\hat \alpha_i$, evaluated for \eqref{Horndeski_simple}, then satisfy
\begin{align} \label{alphaslum}
HM^2\hat{\alpha}_M&= 2 \dot \phi G_{4,\phi}, \quad HM^2\hat{\alpha}_B=2\dot{\phi}\left(XG_{3,X}-G_{4,\phi}\right), 
\end{align}
where the running of the Planck mass $\hat \alpha_M$ is quantified via $HM^2\hat{\alpha}_M\equiv\frac{d}{dt}M^2$. 
Importantly this means that $G_2$ is only implicitly constrained via requiring a $\Lambda{\rm CDM}$ background evolution, i.e. there is no cosmologically relevant explicit dependence on $G_2$ at the level of linear perturbations.

\begin{figure}[t]
\begin{center}
\includegraphics[trim={0.2cm 0.6cm 1.3cm 1.8cm},clip,width=.9\linewidth]{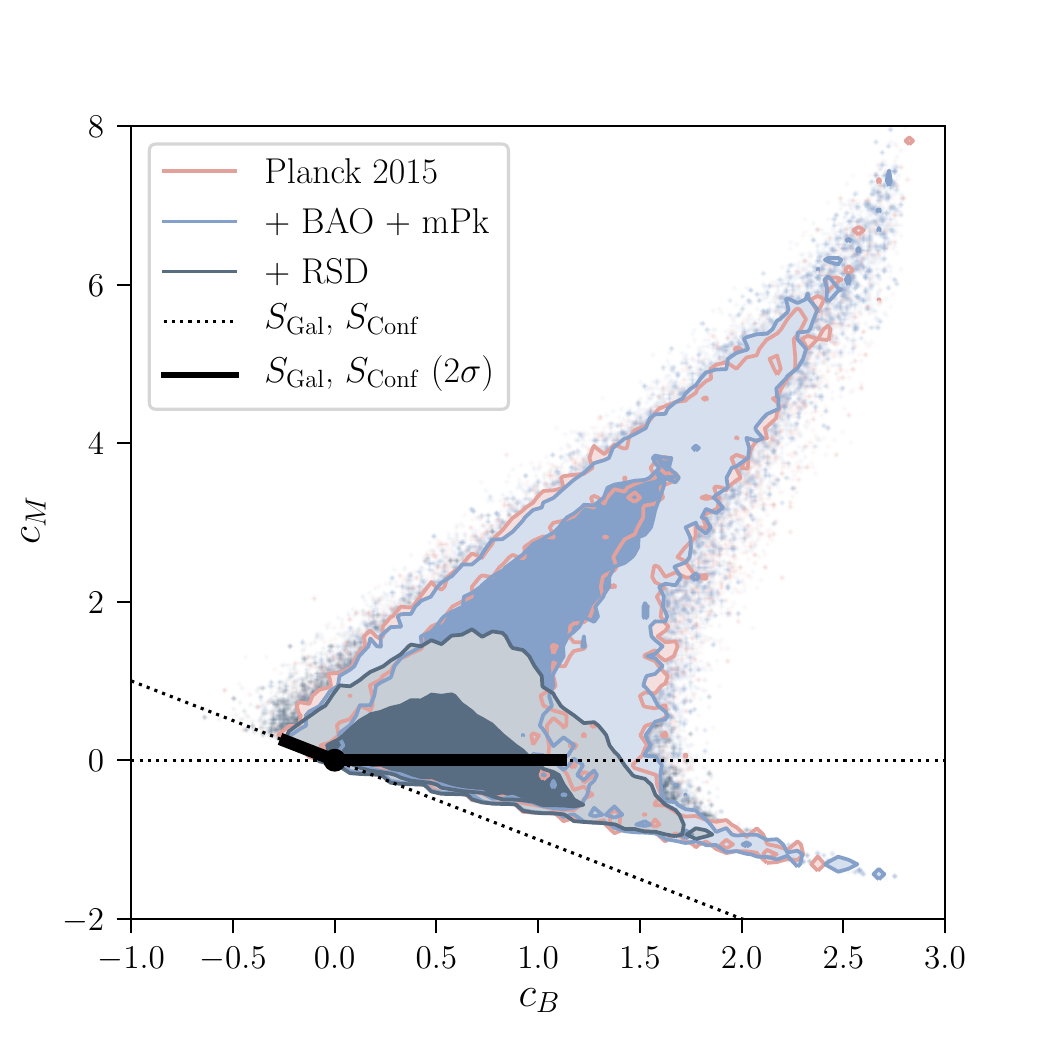}
\end{center}
\caption{Cosmological parameter constraints for the reduced Horndeski theory \eqref{Horndeski_simple} using $\hat \alpha_i = c_i \Omega_{\rm DE}$ \eqref{oParam}. Contours mark $1$ and $2\sigma$ confidence intervals. Adding BAO and mPk data to Planck constraints only has marginal effects,  
whereas adding RSD data significantly improves constraints (mainly by ruling out large positive $c_M$). GR (and all theories with $G_3 = 0$ and $G_{4,\phi} = 0$)
are located at the origin. 
Dotted lines are the regions corresponding to the radiative stability-motivated subsets $S_{\rm Gal}$ \eqref{Horndeski_v1} (horizontal line) and $S_{\rm Conf}$ \eqref{Horndeski_v2} (other line). Thick bars on dotted lines denote the $2\sigma$ confidence region for those theories using Planck + BAO + mPk + RSD data. \label{fig1}}
\end{figure}

Current observations constrain $\hat\alpha_M$ and $\hat\alpha_B$ at the $\Oo$ level (see figure \ref{fig1}), while near-future observations are expected to tighten bounds by approx. one order of magnitude \cite{Alonso:2016suf}. To understand what this implies for interactions in \eqref{Horndeski_simple}, it is instructive to consider the following example 
\begin{align} 
G_2 &= X, &G_3 &= \frac{c_3}{\Lambda_3^3} X, &G_4 &= \frac{M_{\rm Pl}^2}{2}\left(1 + \frac{c_4 \phi^2}{\MPl \Lambda_\star}\right). 
\label{example}
\end{align}
In essence we consider the $G_i$ to be defined via their Taylor expansion in terms of the fields and keep the lowest order non-trivial terms. Note that we have implicitly normalised the zeroth-order piece of $G_4$ and removed its linear dependence on $\phi$ to diagonalise propagators. The $c_i$ are assumed to be $\Oo$ constant coefficients, which amounts to imposing a naturalness assumption.
$\Lambda_3$ and  $\Lambda_\star$ are mass scales (where $\Lambda_3$ is taken to be $\Lambda_3^3 \sim \MPl H_0^2$ as usual) and we will also use the scale $\Lambda_2^4 \equiv \MPl \Lambda_3^3$. 
We can now compute the $\hat\alpha_i$ for \eqref{example}. Defining the dimensionless $\hat\phi \equiv \phi/\Lambda_\star$ and $\hat X \equiv X/\Lambda_2^4$, we find 
\begin{align} \label{alphasEx}
\hat{\alpha}_M &\sim c_4\sqrt{\hat X}\hat \phi, &\hat{\alpha}_B &\sim \sqrt{\hat X}\left(c_3 \hat X - c_4 \hat \phi \right), 
\end{align}
where we 
have ignored overall $\Oo$ numerical factors and
consider late times relevant for dark energy, so $H \sim H_0$. We have also used $\dot\phi^2\sim 2X$ and assumed $G_4 \sim \MPl^2/2$ at leading order (i.e. $\phi^2/(\MPl \Lambda_\star) \ll 1$). 
If the $G_3$ interaction is to have an $\Oo$ effect on the $\hat\alpha_i$ at late times, this implies that at those times $\hat X \sim \Oo$.
If simultaneously $G_4$ has an $\Oo$ effect, this additionally imposes $\hat\phi \sim \Oo$.

Importantly we can check that the above conclusion remains true when also including higher order terms in the expansion of the $G_i$ above, 
so it is not an artefact of the specific example considered above. To show this consider the following generalisation of \eqref{example}
\begin{align} \label{example2}
G_3 &= \frac{c_3}{\Lambda_3^3} \frac{X^{n+1}}{\Lambda_2^{4n}} \frac{\phi^m}{\Lambda_\star^m}, &G_4 &= \frac{M_{\rm Pl}^2}{2}\left(1 + \frac{c_4 \phi^l}{\MPl \Lambda_\star^{l-1}}\right),
\end{align}
where $n,m \geq 0$, $l \geq 2$ and $G_2 = X$ as before. Higher powers of $X$ are suppressed by $\Lambda_2^4$, mimicking known radiatively stable shift-symmetric setups \cite{Pirtskhalava:2015nla} and non-zero $m$ signals shift-symmetry breaking for the $G_3$ interactions, where we choose the symmetry-breaking scale to be $\Lambda_\star$, just as for $G_4$. One may sum over $n,m,l$, but for our purposes zooming in on specific choices of these powers will be sufficient. Computing the $\hat\alpha_i$ as before, we now find
\begin{align} \label{alphasEx2}
\hat{\alpha}_M &\sim c_4\sqrt{\hat X}\hat \phi^{l-1}, &\hat{\alpha}_B &\sim \sqrt{\hat X}\left(c_3 \hat X^{1+n} \hat\phi^{m}- c_4 \hat \phi^{l-1} \right), 
\end{align}
where we ignore $\Oo$ numerical factors (also assuming $l,m,n$ are $\Oo$). As before then, for both $G_3$ and $G_4$ to have $\Oo$ effects on the $\hat\alpha_i$, we generically require $\hat\phi \sim \Oo \sim \hat X$.

Before finishing this section, note that introducing a scale $\Lambda_\star \ll \MPl$ of course also modifies the background equations, in particular in principle also resulting in contributions enhanced by (one power of) $\MPl/\Lambda_\star$. We will leave an investigation of the associated background behaviour for future work and implicitly assume that the freedom in $G_2$ (that does not directly affect the $\hat \alpha_i$ and we will therefore mostly ignore, as discussed above) is sufficient to yield a $\Lambda{}$CDM-like background evolution.
\\

\noindent \textbf{Radiative stability and power counting:} The (classical) predictions of a theory are only trustworthy, if loop corrections are parametrically suppressed for the energy scales one is interested in. If this is not the case, any tree-level computation is swamped by loop effects and (in the absence of a known resummation or UV completion) no reliable prediction can be obtained from the theory. 
In GR, for example, loop corrections are suppressed by powers of the Planck scale, so at energy scales $\Lambda \ll M_{\rm Pl}$ one can reliably extract predictions, whereas at energy scales $\Lambda \gtrsim M_{\rm Pl}$ this is not the case. 
Once one considers departures from GR in a cosmological context, however, the interactions associated with new gravitational degrees of freedom typically bring this cutoff down significantly, so it is no longer trivial to obtain reliable (classical) predictions from such theories. 

To see what this implies for \eqref{Horndeski_simple}, it will be useful to again start by considering the simple example $G_i$ in \eqref{example} and transform to the Einstein frame. 
Doing so yields 
\be
\label{Sloops}
S_E=\int \mathrm{d}^4x \sqrt{-g}\left\{\tfrac{1}{2}\MPl^2 R + X\left(1 - \frac{c_3[\Phi]}{\Lambda_3^3} + \frac{c_4^2\phi^2}{\Lambda_\star^2}\right) 
\right\},
\ee
where the smaller of $\Lambda_3$ and $\Lambda_\star$ becomes the strong coupling scale and we have dropped terms suppressed by powers of $\Lambda_\star/\MPl$ (assuming $\Lambda_\star \ll \MPl$) and absorbed a numerical $\Oo$ factor into $\Lambda_\star$.
The Einstein frame version of the theory now also makes it obvious that $\Lambda_\star$ is the scale associated to (shift) symmetry breaking in this theory. Here we will focus on cases where this scale is significantly below the Planck scale (for a related discussion, see the appendix).
Power-counting and considering pure scalar Feynman diagrams in 4D with $I$ internal legs, $V_1$ $X\hat\phi^2$ and $V_2$ $X[\Phi]$ vertices, we can estimate the types of interactions that will be generated by loops on dimensional grounds and schematically obtain
\be \label{loopscheme}
\frac{1}{\Lambda_\star^{2 V_1}} \frac{1}{\Lambda_3^{3 V_2}} \partial^{4 - 2 V_1 + 2 I} \phi^{4 V_1 + 3 V_2 - 2 I}.
\ee
Focusing on one-loop diagrams with $I = V_1 + V_2$, if only the $c_3$ or $c_4$ vertex are present, 
we obtain loop corrections such as
\begin{align} \label{simpleloops}
\frac{1}{\Lambda_3^{3V_2}} \partial^4[\Phi]^{V_2} \quad \text{and} \quad X \cdot \frac{X}{\Lambda_\star^4} \cdot \hat\phi^{2V_1-4},
\end{align}
respectively. In the $c_3$ case, these corrections are suppressed, if higher-order derivatives are sub-dominant. In the $c_4$ case, these loops are suppressed if $X \ll \Lambda_\star^4$, assuming $\hat\phi\sim\Oo$ and implying $\Lambda_\star \gg \Lambda_2$ (if $\hat X \sim \Oo$). 
Consequently, if only one type of vertex is present, at least these specific loop corrections can remain under control for certain parameter choices. 

If both $c_3$ and $c_4$ interactions are present, however, this picture significantly changes. 
To see how, now again consider the generalised example \eqref{example2} (we will come back to the \eqref{example} in the following section)
%
%
and transform the theory to the Einstein frame. We then obtain the following action 
\be \label{Sloops2}
S_E=\int \mathrm{d}^4x \sqrt{-g}\left\{\tfrac{1}{2}\MPl^2 R + X\left(1 - c_3\hat X^n \hat\phi^m\frac{[\Phi]}{\Lambda_3^3} + c_4^2\hat\phi^p\right) 
\right\},
\ee
again dropping numerical $\Oo$ factors, assuming $\Lambda_\star \ll \MPl$ and where we denote $p = 2 l -2$. Power-counting as before, we estimate loop corrections to have the following form
\be \label{loopscheme2}
\frac{1}{\Lambda_\star^{p V_1+m V_2}} \frac{1}{\Lambda_3^{3 V_2}}  \frac{1}{\Lambda_2^{4 n V_2}}\partial^{4 + (2n + 2) V_2} \phi^{p V_1 + (1+2n+m) V_2},
\ee
where we have already set $I = V_1 + V_2$. The exact way in which derivatives are `distributed' over fields is closely tied to the specific form taken by interaction terms and matters for estimating the size of interactions on a given background. We will show this with several examples below, but in the absence of additional information about the interactions in \eqref{Horndeski_simple} or \eqref{example2}, one would expect all such `distributions' to be generated. Starting with this generic picture in mind and considering choices of $V_1$ and $V_2$ that yield at least as many fields as derivatives, from \eqref{loopscheme2} we would therefore expect to generate terms such as
\begin{align} \label{largeloops2}
X \hat \phi^N \hat X^{(n+1)V_2} \left(\frac{\MPl}{\Lambda_\star}\right)^{V_2}\frac{X}{\Lambda_\star^4}, 
\end{align}
where $N = p V_1 - 4 + (m-1) V_2$ and we note that $\hat X \MPl = X/\Lambda_3^3$. If $\hat X$ is sufficiently small, i.e. if  $\hat X \MPl \ll \Lambda_\star$, these corrections will indeed be suppressed. However, if we assume $\hat X \sim \Oo \sim \hat\phi$ in order to have observable effects on the $\hat\alpha_i$ from {\it both} $G_3$ {\it and} $G_4$ interactions, then for large $V_2$ and positive $N$ the $\MPl/\Lambda_\star$ enhancement in \eqref{largeloops2} wins out over any other potential suppression coming from $X/\Lambda_\star^4$.
%
\footnote{Implicit technical assumptions for this argument are further discussed in the appendix and we re-emphasise that we assume $\Lambda_\star \ll \MPl$ throughout. Note that assuming positive $N$ in the above requires positive $m$ and/or $p$. Finally it is worth stressing that the problematic behaviour in \eqref{largeloops2} is primarily associated with the $V_2$ vertices and corrections of the type shown above are possible (also in the absence of any $V_1$ interactions) for $m \geq 2$.}  
Since loop corrections for all choices of $V_1$, $V_2$ are generated, one would therefore generically expect these corrections to dominate over the classical ansatz, rendering the theory unpredictive.
%
%
\\

\noindent {\bf Shift-symmetric vs. symmetry-breaking $G_3$ interactions}: 
How robust is the `generic' result \eqref{largeloops2}? An obvious strategy to improve the radiative properties of \eqref{Horndeski_simple} would be to endow it with more symmetry, but other than the general Galilean or shift symmetric cases (that eliminate any non-trivial $G_4$ alltogether), it is not obvious what other sensible choices exist in this regard. However, as we shall see, the case when just $G_3$ is endowed with a Galilean or shift symmetry has a number of attractive features, even if a symmetry-breaking $G_4$ is present, i.e. despite the fact that these symmetries are then not respected by the theory as a whole. 
To see why, recall \eqref{Sloops2}, 
%
%
and first focus on the case of $m=0$ i.e. a setup with a shift-symmetric $G_3$. We will choose $p=2$ in what follows for simplicity, but the specific power of $p$ (i.e. the precise form of $G_4$) will not be important for the argument.
We now follow the argument of \cite{Luty:2003vm} and schematically consider the contribution to a one-loop Feynman diagram from the above $c_3$ vertex. The most dangerous interactions are those, where the smallest number of derivatives acts on the external legs. Schematically labelling external and internal legs with `e' and `i' indices respectively, a potentially dangerous interaction would therefore be
\begin{align} \label{NR1}
\partial_{\mu_1} \phi_{\rm i} \partial^{\mu_1} \phi_{\rm e} \ldots \partial_{\mu_{n+1}} \phi_{\rm e} \partial^{\mu_{n+1}} \phi_{\rm e} \Box \phi_{\rm i},
\end{align}
where all external legs carry just one derivative. However, we can re-write this as
\begin{align} \label{NR2}
\partial^{\mu_1} \phi_{\rm e} \ldots \partial_{\mu_{n+1}} \phi_{\rm e} \partial^{\mu_{n+1}} \phi_{\rm e}
\partial_\alpha \Big[ \partial^\alpha\phi_{\rm i} \partial_{\mu_1} \phi_{\rm i} - \tfrac{1}{2} \delta_{\mu_1}^\alpha \partial_\beta \phi_i \partial^\beta \phi_i\Big].
\end{align}
Written in this form, it becomes obvious that, after integration-by-parts, one external leg always carries two derivatives (while the others carry one, as before). With the knowledge that each $V_2$ vertex in the shift-symmetric $G_3$ case therefore contributes one external leg with two derivatives, we then obtain loop corrections such as\footnote{Note that we need at least two $V_1$ vertices to generate loop corrections of this particular form.}    
%
%
\begin{align} \label{Sloops3}
(\partial \phi)^2 \cdot \left(\frac{\partial^2\phi}{\Lambda_3^3}\right)^{V_2} \cdot \left(\frac{(\partial \phi)^2}{\Lambda_2^4}\right)^{n V_2} \cdot \frac{\partial^2}{\Lambda_\star^2} \cdot \hat\phi^{2V_1-2}
\end{align}
The second and third term give order one contributions on cosmological backgrounds with $\partial^2\phi_0 \sim \Lambda_3^3$ and $\partial\phi \sim \Lambda_2^2$, but the $\partial^2/\Lambda_\star^2$ operator provides additional suppression, as long as $\Lambda_\star \gg \Lambda_2$ (as assumed throughout). 
Dangerous loop interactions of the form in \eqref{largeloops2} are therefore not present in this case. Indeed, on cosmological backgrounds with $\phi_0 \sim \Lambda_\star$ and $X_0 \sim \Lambda_2^4$, the above interaction will be suppressed by $(\Lambda_2/\Lambda_\star)^4$, so will be sub-dominant as long as $\Lambda_\star \gg \Lambda_2$ as before.
For a shift-symmetric $G_3$ case the general theory \eqref{Horndeski_simple} is therefore still endowed  with parametrically suppressed radiative corrections -- a remarkable property, given the symmetry breaking nature of the theory as a whole.

The above argument is highly reminiscent (and a generalisation) of the non-renormalisation theorem for the pure Galileon \cite{Luty:2003vm}. This follows from \eqref{NR1} and \eqref{NR2} in the case where \eqref{NR1} is just $\partial_\mu \phi_i \partial^\mu\phi_e \Box\phi_i$ and for pure Galileons implies that {\it all} external legs come with at least two derivatives, so Galileon loops do not re-generate the standard Galileon interactions (since these have less than two derivatives per field). When we now switch on a non-trivial $G_4$ in \eqref{Sloops2} in addition to having a Galileon $G_3$ interaction, i.e. $n=0, m=0$, the above argument straightforwardly shows that a remnant of this non-renormalisation theorem also still applies in this special case. Considering a loop with $V_2$ cubic Galileon $c_3$ vertices and $V_1$ (Galilean and shift) symmetry-breaking $c_4$ vertices (where $V_1 \geq 2$ as before), we then expect to generate loop corrections of the form 
\begin{align} \label{galloops1}
(\partial \phi)^2 \cdot \left(\frac{\partial^2\phi}{\Lambda_3^3}\right)^{V_2} \cdot \frac{\partial^2}{\Lambda_\star^2} \cdot \hat\phi^{2V_1-2},
\end{align}
which are suppressed by the same argument as for \eqref{Sloops3}. The only difference of this special case is that there is only one external leg per $V_2$ vertex and so no external legs with less than two derivatives attach to a $V_2$ vertex here.

Following the more symmetric cases considered above, we can now finally turn to 
an explicit example of a theory with a Galileon and shift symmetry breaking $G_3$ interaction. We consider \eqref{Sloops2} with $m=2, n=0, p=2$.\footnote{Note that, upon integrating-by-parts, the $\hat\phi^2 \cdot X \cdot [\Phi]/\Lambda_3^3$ interaction then gives rise to (among other terms) an interaction that goes as $\hat\phi \cdot X \cdot X/(\Lambda_\star \Lambda_3^3)$. This is already a sign of problematic behaviour, since higher powers of $X$ are then no longer suppressed by the scale $\Lambda_2$, calling into question the validity of the cosmological background solution. As we shall see below, this also signals the onset of problematic loop corrections in the present case.} 
Nothing now prevents all external legs attached to $V_2$ vertices in one-loop Feynman diagrams to carry at most one derivative in this setup. We can see this explicitly with a quick example. Consider the following contribution to a one-loop Feynman diagram from a shift-symmetry breaking $\phi (\partial \phi)^2 \Box \phi$ interaction 
\begin{align} \label{NR1-2}
\phi_{\rm e} \partial_{\mu} \phi_{\rm i} \partial^{\mu} \phi_{\rm e} \Box \phi_{\rm i} = 
\phi_{\rm e} \partial^{\mu} \phi_{\rm e} \partial_\alpha \Big[ \partial^\alpha\phi_{\rm i} \partial_{\mu} \phi_{\rm i} - \tfrac{1}{2} \delta_{\mu}^\alpha \partial_\beta \phi_i \partial^\beta \phi_i\Big],
\end{align}
where we have already applied the re-writing employed above in going from \eqref{NR1} to \eqref{NR2}. Unlike before, this no longer guarantees that, up to integration-by-parts, we can re-write this contribution with two derivatives acting on an external leg. Instead we now also generate $G_2$ type interaction terms in this way, where only at most one derivative acts on all the fields involved.
More specifically, we now both generate interaction terms, where one external field carries two derivatives (while all others carry zero or one), {\it as well as} the above-mentioned $G_2$ type interaction terms, where one power of $X$ is suppressed by $\Lambda_\star \Lambda_3^3$ instead of $\Lambda_2^4$. This is related to the discussion of enhanced contributions to the background equations of motion below \eqref{alphasEx2}. A potential issue can therefore already be spotted here before investigating loop corrections, but nevertheless considering these corrections offers a potent way of diagnosing a problem with such theories.
Doing so we can therefore now estimate loop corrections to e.g. take the following form 
\begin{align} \label{mixEx}
(\partial \phi)^2 \cdot \left(\frac{(\partial \phi)^2}{\Lambda_\star \Lambda_3^3}\right)^{V_2} \cdot \frac{\partial^2}{\Lambda_\star^2} \cdot \hat\phi^{2V_1+V_2-2}.
\end{align}
While the $\partial^2/\Lambda_\star^2$ operator provides a suppressing effect, as in the previous example, the second term can be re-written as $\hat X^{V_2} \cdot (\MPl/\Lambda_\star)^{V_2}$, where the $(\MPl/\Lambda_\star)^{V_2}$ factor could provide an arbitrarily large enhancement as long as $\MPl \gg \Lambda_\star$ (as assumed throughout). In this setup we therefore expect to lose radiative stability and with it control of the theory.\footnote{Note that the symmetry breaking scale $\Lambda_\star$ entering into $G_3$ was crucial to this argument -- radiative stability is restored if $\phi$ is suppressed by $\MPl$ in the $G_3$ interaction, but it then no longer meaningfully contributes to the $\hat\alpha_i$ in the setup we are considering, so we relegate this case to the appendix.} 

While shift-symmetric (and Galilean) $G_3$ therefore allow \eqref{Horndeski_simple} to still enjoy protection from radiative corrections, theories with a shift-symmetry breaking $G_3$ (and hence explicit $\phi$-dependence) do not generically share that property.
\\

\noindent{\bf Two well-motivated radiatively stable theories}:
A key conclusion following from the above argument then is that, as long as $\Lambda_\star \ll \MPl$, there is some tension for fully generic choices of $G_3$ and $G_4$ in \eqref{example2} to {\it both} yield observable contributions to the $\hat\alpha_i$ as well as to do so in a controllable (radiatively stable) fashion. Nevertheless, several well-defined (sub-)classes of theories where this is possible do exist.
%
%
%
%
For the remainder of this paper we therefore want to focus on two particularly well-motivated such classes: First the case where the whole theory is endowed with a shift symmetry (eliminating non-trivial $G_4)$ and secondly the case where no symmetry requirement is imposed on the whole theory and the dominant operator for linear perturbations is the shift-symmetry breaking $G_4$ term, while the higher derivative operators associated to $G_3$ are suppressed. Both of these cases are well-motivated from first principles and do not require tuning interactions to have separate symmetries (for different subsets of interactions). For the first case we have
\begin{align} \label{Horndeski_v1}
S_{\rm Gal} =\int \mathrm{d}^4x \sqrt{-g}\left\{ G_2(\phi,X) -G_3(X)[\Phi] + \tfrac{1}{2}M_{\rm Pl}^2R\right\},
\end{align}
where $G_4$ is constant and has been normalised to $\tfrac{1}{2}M_{\rm Pl}^2$ and the $\phi$-dependence in $G_3$ is eliminated by the $\phi \to \phi + c$ shift symmetry. Terms linear/quadratic in $\phi$ are protected by their own renormalisation theorems \cite{Porrati:2004yi,Nicolis:2004qq,Endlich:2010zj,GalInf}, and hence at least for the covariant Galileon \cite{Deffayet:2009wt} only `softly' break this shift symmetry \cite{GalInf}, which is why we have kept $\phi$-dependence for $G_2$. Modulo this caveat, \eqref{Horndeski_v1} is therefore a subset of weakly broken Galilean (WBG) theories \cite{Pirtskhalava:2015nla}, with loop corrections parametrically suppressed as a result of the (weakly broken) Galileon symmetry \cite{Nicolis:2008in,Luty:2003vm}. 
We can see this explicitly for the example discussed above. Using \eqref{loopscheme2}, $S_{\rm Gal}$ corresponds to setting $V_1 = 0 = m$. There are therefore always more derivatives than fields and (for large $V_2$ and $n \geq 1$) we expect corrections such as 
\begin{align}
X \hat X^{nV_2 - 3} \left(\frac{[\Phi]}{\Lambda_3^3}\right)^{V_2}\left(\frac{[\Phi]}{\Lambda_2^3}\right)^{4}.
\end{align}
For backgrounds with $\hat X \sim \Oo \sim [\Phi]/\Lambda_3^3$, as considered by \cite{Pirtskhalava:2015nla}, the final factor of $[\Phi]/\Lambda_2^3$ then suppresses these corrections (since $\Lambda_2 \gg \Lambda_3$). 
%

%
The second well-motivated subset of \eqref{Horndeski_simple} we consider is 
\begin{align} \label{Horndeski_v2}
S_{\rm Conf} =\int \mathrm{d}^4x \sqrt{-g}\left\{ G_2(\phi,X) + G_4(\phi) R\right\},
\end{align}
where $G_3$ is now absent, but the conformal $G_4(\phi)$ coupling to $R$ is kept. 
Using \eqref{loopscheme2}, this case corresponds to setting $V_2 = 0$ in the example above and hence we there expect corrections such as 
\begin{align}
X \frac{X}{\Lambda_\star^4} \hat\phi^{pV_1-4},
\end{align}
where the $X/\Lambda_\star^4$ factor suppresses these loop-generated interactions on backgrounds where $\hat X \sim \Oo$ and as long as $\Lambda_\star \gg \Lambda_2$ (cf. the discussion around \eqref{simpleloops}). The $S_{\rm Conf}$ class of theories can likely be reduced further with radiative stability arguments, 
but here we simply note that it also contains known radiatively stable models, especially ones arising as low-energy EFTs from dimensionally reduced higher-dimensional constructions (see \cite{Clifton:2011jh} for a review). Arguably the prime example here are JBD theories \cite{Brans:1961sx}, that arise as a low-energy EFT in Randall-Sundrum I setups \cite{Randall:1999ee}.
\\

\noindent {\textbf{Cosmological parameter constraints:} 
We now want to extract the cosmological predictions for the well-motivated subsets \eqref{Horndeski_v1} and \eqref{Horndeski_v2} and contrast them with the predictions from the general \eqref{Horndeski_simple}. The linear cosmology as parametrised by \eqref{alphaslum} significantly simplifies for \eqref{Horndeski_v1} and \eqref{Horndeski_v2} and we find
\begin{align}
{S_{\rm Gal}} \quad &\Rightarrow \quad \hat{\alpha}_M= 0 \quad \text{and} \quad \hat{\alpha}_B=\frac{2 \dot{\phi}X}{HM^2}G_{3,X},\nonumber \\
S_{\rm Conf} \quad  &\Rightarrow \quad \hat{\alpha}_M= -\hat\alpha_B = \frac{2\dot \phi}{HM^2} G_{4,\phi}.
\label{radalphas}
\end{align}
This means that these theories effectively give rise to linearised cosmologies that are described by just {\it one} free function: $\hat \alpha_B$ (once a background has been specified, i.e. the evolution of $H$ has been given). Note that the form of $\hat \alpha_B$ is strongly constrained, since $G_4$ is a function of $\phi$ only and $G_3$ is a function of $X$ only, for ${S}_{\rm Conf}$ and ${S}_{\rm Gal}$ respectively.

\begin{figure}[t]
\begin{center}
\includegraphics[trim={0.5cm 0.7cm 0.5cm 0.5cm},clip,width=0.9\linewidth]{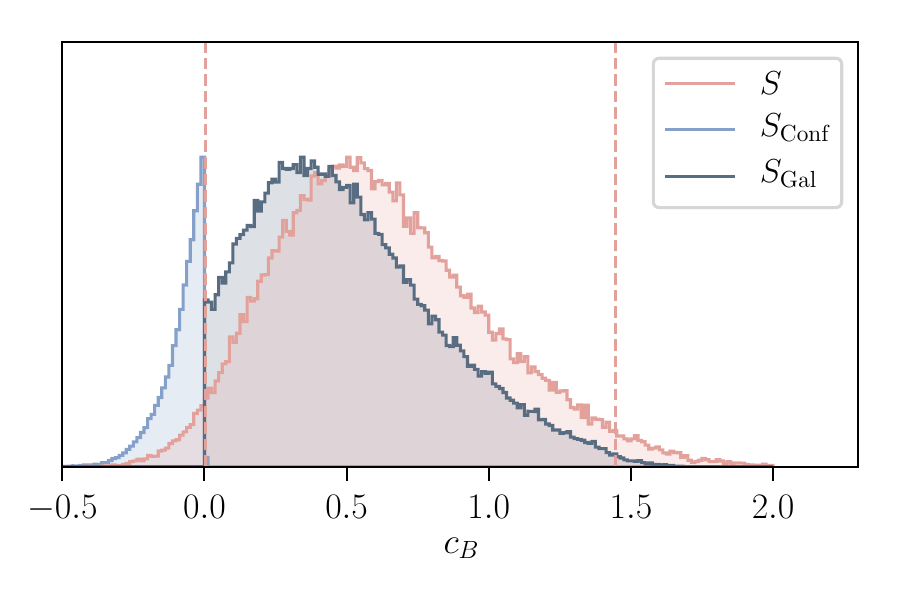}
\end{center}
\caption{Here we plot the 1D posterior distribution of $c_B$, corresponding to the Planck + BAO + mPk + RSD contours from Figure 1. 
$c_B$ is shown for the reduced Horndeski theory \eqref{Horndeski_simple} and its well-motivated and radiatively stable subsets $S_{\rm Gal}$ and $S_{\rm Conf}$. Note that $c_B$ is the only free modified gravity parameter for $S_{\rm Gal}$ and $S_{\rm Conf}$ and that the regions populated by $S_{\rm Gal}$ and $S_{\rm Conf}$ are mutually exclusive. Dotted vertical lines denote $2\sigma$ bounds for $S$, showing that applying the theoretical prior $S_{\rm Conf}$ pushes one outside the $2\sigma$ region for $S$. 
We have normed histograms to have the same `maxima'.
\label{fig2}}
\end{figure}

To demonstrate the strength of the above constraints more quantitatively, we now ought to pick a parametrisation for the  key functions $\hat\alpha_M,\hat\alpha_B$ in $S$ \eqref{Horndeski_simple} vs. just $\hat \alpha_B$ as identified in \eqref{radalphas}. Numerous such parametrisations exist -- for a discussion of relative merits see \cite{Bellini:2014fua,BelliniParam,Linder:2015rcz,Linder:2016wqw,Gleyzes:2017kpi,
Alonso:2016suf,ourEnc} and previous related constraints include \cite{BelliniParam,Kreisch:2017uet}. 
For simplicity here we will pick arguably the one most frequently used \cite{Bellini:2014fua}: 
\be \label{oParam}
\hat\alpha_i = c_i \Omega_{\rm DE}.
\ee 
This parametrises each $\hat\alpha_i$ in terms of just one extra constant parameter $c_i$ and the proportionality to the fractional contribution of dark energy to the energy density of the universe, $\Omega_{\rm DE}$, ensures that the modification is a late time effect. This parametrisation is known to accurately capture the evolution of a wide sub-class of Horndeski theories \cite{Pujolas:2011he,Barreira:2014jha}, but not all \cite{Linder:2016wqw}. While therefore undoubtedly in need of future refinement, it provides an excellent quantitative illustration of the constraining power of the results derived above.

We now perform a Monte Carlo Markov Chain (MCMC) analysis, using Planck 2015 CMB (cosmic microwave background) temperature, CMB lensing and low-$\ell$ polarisation data \cite{Planck-Collaboration:2016af, Planck-Collaboration:2016aa, Planck-Collaboration:2016ae}, BAO (baryon acoustic oscillation) measurements from SDSS/BOSS \cite{Anderson:2014, Ross:2015}, constraints from the SDSS DR4 LRG matter power spectrum (mPk) shape \cite{Tegmark:2006} and RSD (redshift space distortions) constraints from BOSS and 6dF \cite{Beutler:2012, Samushia:2014}.
We compute constraints on $c_{B}$ and $c_{M}$, marginalising over the standard $\Lambda{\rm CDM}$ parameters $\Omega_{\rm cdm}, \Omega_{\rm b}, \theta_s,A_s,n_s$ and $\tau_{\rm reio}$. 
Note that we infer $\Omega_{\rm DE}$ through the closure equation (assuming no cosmological curvature).
Furthermore, we impose that the asymptotic value of the effective Planck mass $M$ at early times is indeed $M_{\rm Pl}$, since we do not wish to constrain early universe modifications of gravity here (for a different approach see \cite{BelliniParam}). 
We put no priors on the absence of `classical' (e.g. gradient) instabilities, 
since we find that the constraints derived with and without such priors are near-identical. In a nutshell: the data will exclude any model with a significant such instability. 
The only difference are small regions in parameter space, where a hard prior would erroneously exclude models that display transient instabilities during radiation domination, which are an artefact of the choice of parametrisation and do {\it not} affect observables. 
For additional details regarding the MCMC implementation see \cite{ourEnc}.\footnote{We note that, while this work was being finalized, the Planck Collaboration published its final results. The most significant difference between these and the 2015 results used here is the shift to a lower value of the optical depth to reionization, $\tau_{\rm reio}$, by approximately $1.5\sigma$. As there are no strong correlations between the value of this value and the $\alpha_i/c_i$ parameters discussed here, we believe that these new constraints will not significantly affect our conclusions. Secondly, note that the inclusion of additional RSD likelihoods would likely yield stronger constraints. We prefer to remain conservative here and do not mix constraints from RSD measurements with similar redshifts, but leave a proper analysis of the impact of adding additional RSD or other large scale structure likelihoods (while carefully accounting for potential cross-correlations) for the future.}

Figure \ref{fig1} shows the cosmological constraints on the modified gravity parameters: $c_M$ and $c_B$. BAO and mPk data only mildly modify the Planck constraints, but RSD measurements significantly tighten constraints, especially on $c_M$. 
This is rather intuitive, as RSDs constrain $f\sigma_8$, which is particularly sensitive to the strength of gravity as measured by the effective Planck mass. It is worth emphasising that the lower (small and negative $c_M$) border of the contours in figure \ref{fig1} is determined by the onset of strong gradient instabilities (illustrated by the absence of accepted points in that region). These instabilities occur when the scalar speed of sound $c_s$ becomes imaginary, specifically 
\be \label{gradient_condition}
c_s^2 = \frac{(2-\hat\alpha_B)\left(\tfrac{1}{2}\hat\alpha_B + \hat\alpha_M\right) +\frac{2 \dot H}{H^2}\left(\frac{1-M^2}{M^2}\right) + \frac{\tfrac{d}{dt}{(\hat\alpha_B H)}}{H^2}}{\hat\alpha_K + \tfrac{3}{2}\hat\alpha_B^2} < 0, 
\ee
where the expression given assumes a $\Lambda{}{\rm CDM}$ background. The well-motivated models \eqref{Horndeski_v1} and \eqref{Horndeski_v2} 
%
trace out two lines in figure \ref{fig1}: $c_M = 0$ for $S_{\rm Gal}$ and $c_M = -c_B$ for $S_{\rm Conf}$. 
Figure \ref{fig2} illustrates the 1D posterior distribution for $c_B$, showing that the sole remaining modified gravity parameter (at the level of linear perturbations) is tightly constrained. Specifically, we find the following bounds 
\begin{align}
{S_{\rm Gal}} \quad &\Rightarrow \quad c_B > 0 \quad \text{and} \quad c_B <  1.11 \;\; (2\sigma),\nonumber \\
{S_{\rm Conf}} \quad &\Rightarrow \quad c_B < 0 \quad \text{and} \quad c_B >  -0.24 \;\; (2\sigma),\nonumber \\
S \quad  &\Rightarrow \quad 0.01 < c_B <  1.45 \;\; (2\sigma).
\label{Ssigmas}
\end{align}
Gradient stability constraints enforce that $c_B > 0$ for ${S_{\rm Gal}}$ and $c_B < 0$ for ${S_{\rm Conf}}$, i.e. they ensure the two models occupy mutually exclusive parts of parameter space. 
Note that the $c_B$ constraints for ${S_{\rm Gal}}$ and ${S_{\rm Conf}}$ are virtually unaffected by the addition of further data to Planck measurements, so unlike for $S$, RSDs here do not provide significant additional constraining power.
Interestingly the region populated by $S_{\rm Conf}$ would have been excluded at $2\sigma$ when using $S$ as the fiducial model. 
%
This illustrates the strong impact theoretical priors can have. 
A further interesting observation that follows from the findings presented here is the following: If future observations drive observational constraints in the $c_B - c_M$ plane into regions consistent with $S$ and $\Oo$ $\hat\alpha_i$, but not overlapping with ${S_{\rm Gal}}$ or ${S_{\rm Conf}}$, then within the context of \eqref{Horndeski_simple} with $\Lambda_\star \ll \MPl$ this would suggest a theory with a shift- or Galilean-symmetric $G_3$ as well as a shift-symmetry breaking $G_4$ as a natural explanation of the data.\footnote{Note, however, that the pure cosmological ${S_{\rm Conf}}$ constraints computed here do not rely on any assumption for $\Lambda_\star$ themselves.}
%
%
Finally figure \ref{fig2} demonstrates that restricting to the $S_{\rm Gal}$ and $S_{\rm Conf}$ subsets {\it both} reduces the number of free functions {\it and} leads to tighter constraints on the remaining parameters. 
\\

\noindent\textbf{Conclusions:} We can summarise our key results as follows.  
\begin{itemize}[leftmargin=.3cm]
\item Radiative stability: Imposing $c_{\rm GW} = c$ reduces Horndeski ST theories to \eqref{Horndeski_simple}, a theory controlled by the three functions $G_{2,3,4}$.
Power counting arguments show that, for fully generic models, there is a tension between the requirement of radiative stability and {\it both} the conformal coupling to gravity $G_4(\phi)R$ {\it and} the higher-derivative $G_3(\phi,X)[\Phi]$ interaction simultaneously contributing to cosmologically observable deviations from GR,
at least as long as the symmetry breaking scale associated to $G_4$ is significantly below the Planck mass. 
%
However, inspecting the generic power counting argument more closely, we have also identified a number of cases that can give rise to departures from GR within the reach of current and near-future experiments in a radiatively stable manner: Theories with a shift-symmetric $G_3$. Crucially this also remains true in the presence of a symmetry breaking $G_4$ interaction. More generally, there are a number of caveats to the generic argument we have discussed explicitly, so additional radiatively stable subclasses may be found by making use of those caveats.
%
%
%
\item Linear cosmology: Focusing on two particularly well-motivated subsets of such radiatively stable theories, namely shift symmetric theories \eqref{Horndeski_v1} (with a trivial, constant $G_4$) and theories with a conformal coupling of the Horndeski scalar to gravity \eqref{Horndeski_v2} (without a sizeable $G_3$ contribution), we have computed their linear cosmologies. Once the background evolution (and hence $H$) is specified for these subsets of theories, their phenomenology is controlled by just {\it one} additional, free function of time: $\hat\alpha_B$.   
\item Cosmological parameter constraints: Observational constraints on modified gravity parameters are shown in figures \ref{fig1} and \ref{fig2}, with RSDs proving particularly constraining for \eqref{Horndeski_simple}, while constraints for the $S_{\rm Gal}$ and $S_{\rm Conf}$ subsets (\eqref{Horndeski_v1} and \eqref{Horndeski_v2}, respectively) are driven by Planck data and gradient stability conditions.  Interestingly, constraints from gradient instabilities also ensure that $S_{\rm Gal}$ and $S_{\rm Conf}$ occupy mutually exclusive regions in parameter space, offering a promising target to discriminate between these theories in the future. Within the context of the theories explored here, this argument also provides regions in parameter space that will act as a smoking gun signature for the joint presence of shift-symmetric $G_3$ and non-trivial $G_4$ interactions, should future observations drive constraints into these regions.
\end{itemize}
Various extensions of this work suggest themselves, 
with the inclusion of additional datasets holding particular promise (we will discuss the impact of  Galaxy-ISW cross-correlations in \cite{ourISW} -- also see \cite{Renk:2016olm,Renk:2017rzu,Lombriser:2016yzn}). 
%
It would also be very interesting to understand in what precise circumstances a $c_{\rm GW} = c$ tuning is radiatively stable and whether one can use this to impose additional constraints. 
Here we have shown that combining data-driven cosmological parameter estimation with the theoretical requirement of radiative stability, two aspects typically considered separately, can be used to significantly improve the (observational and theoretical) bounds we can place on modified gravity/dark energy models. We hope that the approach outlined here will contribute towards holistically constraining cosmological deviations from GR in the future, taking into account a wide range of observational and theoretical constraints in an integrated manner.

\vspace{-0.1in}

\section*{Acknowledgments}
\vspace{-0.1in}
\noindent We especially thank E. Bellini and S. Melville for numerous discussions and shared insights. We also acknowledge several useful discussions with T. Brinckmann, C. de Rham, P. Ferreira, A. Refregier, A. Tolley and E. Trincherini, as well as feedback from an anonymous referee. JN acknowledges support from Dr. Max R\"ossler, the Walter Haefner Foundation and the ETH Zurich Foundation. AN acknowledges support from SNF grant 200021\_169130. In deriving the results of this paper, we have used: CLASS \cite{Blas:2011rf},  corner \cite{corner}, hi\_class \cite{Zumalacarregui:2016pph}, MontePyton \cite{Audren:2012wb,Brinckmann:2018cvx} and xAct \cite{xAct}.

\appendix
\section*{Appendix} \label{appendix}

\noindent {\bf Two shift-symmetry breaking scales}:
Whenever we have considered cases, where shift-symmetry was broken by the $G_3$ interaction as well, i.e. not just by $G_4$, we have assumed that the associated symmetry-breaking scale is the same for $G_3$ and $G_4$: $\Lambda_\star$, which satisfies $\MPl \gg\Lambda_\star \gg \Lambda_2$. One may be interested in instead considering the following case
\begin{align} \label{example3}
G_3 &= \frac{c_3}{\Lambda_3^3} \frac{X^{n+1}}{\Lambda_2^{4n}} \frac{\phi^m}{\MPl^m}, &G_4 &= \frac{M_{\rm Pl}^2}{2}\left(1 + \frac{c_4 \phi^l}{\MPl \Lambda_\star^{l-1}}\right),
\end{align}
where the shift-symmetry breaking scale for $G_3$ is $\MPl$, while it remains $\Lambda_\star$ for $G_4$. This setup at first sight has promising properties, as the following example shows. Consider the above case with $m=1, n=0, p=2$ (we recall that $p = 2l - 2$).  One then obtain radiative corrections such as
\begin{align}
(\partial \phi)^2 \cdot \left(\frac{(\partial \phi)^2}{\Lambda_2^4}\right)^{V_2} \cdot \hat\phi^{2V_1-4} \cdot \frac{(\partial \phi)^2}{\Lambda_\star^4}.
\end{align}
As long as $\Lambda_\star \gg \Lambda_2$, as indeed we have assumed throughout most of this paper, the last term ensures this contribution is suppressed. However, while this mixed symmetry-breaking scale case is therefore promising from the radiative stability perspective, the contribution of the $G_3$ term to $\hat\alpha_B$ is now suppressed by $(\Lambda_\star/\MPl)^m$. So for the setup considered here, where $\hat\phi \sim \Oo \sim \hat X$, this means a $G_3$ as in \eqref{example3} will only yield a highly suppressed contribution to the $\hat\alpha_i$. As far as linear cosmology is concerned, to leading order we can then drop $G_3$ for this theory altogether and this case reduces to $S_{\rm Conf}$ \eqref{Horndeski_v2}.
\\

\noindent {\bf Loop corrections and internal graviton lines}: When estimating one-loop corrections, we have so far only considered pure scalar interactions, so one may wonder whether loop corrections with internal graviton lines contribute at the same order. Taking \eqref{Sloops} as a simple example and expanding to linear order in the graviton perturbation $h_{\mu\nu}$, where $g_{\mu\nu} = \eta_{\mu\nu} + h_{\mu\nu}$, we find the following interactions (at linear order in $h_{\mu\nu}$)
\begin{align} \label{Lh}
{\cal L}_{\rm h \; lin.} = \hat{h} \left(\tfrac{1}{2} X - c_3 \phi^{\mu} \phi^{\nu} \hat\Phi_{\mu\nu} + \tfrac{1}{2} c_4 X \hat\phi^2 \right),
\end{align}
where we defined $\hat h \equiv h^\mu_\mu/M_{\rm Pl}$ and $\hat \Phi_{\mu\nu} \equiv \Phi_{\mu\nu}/\Lambda_3^3$. Note that the $c_3$-dependent piece is a combination of the contribution that picks up a power of $h$ from the metric determinant and the contribution that picks up one power of $h$ from the (covariant) box operator. In other words, what is important to keep in mind is that
\begin{align}
\nabla^\mu \nabla_\mu \phi = \Box\phi + \Gamma^{\nu}{}_{\mu\nu} \partial^{\mu}\phi = \Box \phi + \tfrac{1}{2}\left(\partial_\mu h_\nu^\nu \right)\partial^\mu \phi,
\end{align}
where $\Gamma$ is the Christoffel symbol for $g_{\mu\nu}$, we define $\Box \equiv \partial_\mu \partial^\mu$ and we integrate-by-parts to remove the derivative from $h$ in obtaining \eqref{Lh}.
The crucial point here is that all $h$-dependent interactions generated from \eqref{Sloops} are $\MPl$ suppressed and hence do not contribute at leading order. Indeed this is not an artefact of linearising in $h$ and it remains true at higher orders as well. All non-linear, $h$-dependent interactions are therefore suppressed in the same way higher order pieces coming from the Einstein-Hilbert term are.\footnote{In other words, at leading order in $\MPl$, the only $h$-dependent piece that remains is therefore the linearised Einstein-Hilbert term as usual.} Note that this becomes more subtle for more general Horndeski theories with gravitational waves not propagating at the speed of light, e.g. with $G_{4,X} \neq 0$. In such cases, i.e. when $G_4$ and/or $G_5$ carry $X$-dependence, non-trivial interactions that are not $\MPl$-suppressed and involve both gravitons and scalars can arise (see e.g. the related discussion in \cite{Pirtskhalava:2015nla}). 

When moving from the simple example \eqref{Sloops} to the more general \eqref{Sloops2}, the above conclusions remain true, except there now is one additional type of interaction, linear in $h$ but no longer Planck suppressed, that is generated whenever $m \geq 1$. Namely we find
\begin{align} \label{Lh2}
{\cal L}_{\rm h \; lin.} = m c_3 X \hat\phi^{m-1} {\hat X}^{n+1} \frac{h_\mu^\mu}{\Lambda_\star} + \ldots,
\end{align}
where the ellipsis contains the interaction terms as in \eqref{Lh} as well as other Planck suppressed terms.
This interaction is only $\Lambda_\star$ suppressed, so can contribute at leading order unless $\Lambda_\star$ is sufficiently large. Note, however, that this should not be viewed as a new restriction on $\Lambda_\star$, since $c_3$ interactions with $m \geq 1$ already generically give rise to dangerous loop corrections just from pure scalar interactions, which dominate over the classical ansatz (as discussed in the main text). So no additional interactions are ruled out when considering loop corrections involving interaction vertices of the form \eqref{Lh2} (with internal graviton lines), even when $h_\mu^\mu/\Lambda_\star \gtrsim {\cal O}(1)$, since the same interactions are already generically disqualified when just considering pure scalar loops.

In summary, interactions involving one or more factors of $h$ derived from \eqref{Sloops} or \eqref{Sloops2} are $\MPl$-suppressed, so loops generated with such interactions (e.g. pure scalar interactions generated via loop diagrams with one or more internal gravitons) will not contribute at leading order. The only exception are loop corrections generated involving interactions of the type \eqref{Sloops2} for $m \geq 1$. However, these are already generically ruled out by considering pure scalar loops, so no additional constraints arise from loop corrections involving internal gravitons for \eqref{Horndeski_simple} here.
\\

\noindent {\bf Caveats}: In addition to the importance of tracking how derivatives are `distributed' over fields in estimating loop corrections, as discussed in the main text, here we want to highlight a few additional implicit assumptions in the argument leading to \eqref{loopscheme2} and \eqref{largeloops2} in the main text. I) Naturalness was a crucial ingredient in our argument, ensuring that $c_i \sim \Oo$. Any physical mechanism (e.g. a new symmetry for specific theories with observationally relevant $G_4$ and $G_3$) that allows stably tuning the $c_i$ may therefore alter the argument. Note, however, that the link between radiative stability and symmetries is rather subtle \cite{deRham:2014wfa}.
II) We have assumed that $\Lambda_\star \ll \MPl$, i.e. that pure graviton interactions are still normalised by $\MPl$ at leading order and we can consistently truncate interactions as in \eqref{Sloops}. Otherwise there is no sensible expansion of $G_4$ in powers of $\phi$, since when $\Lambda_\star \sim \MPl$, $\phi/\MPl$ needs to be $\Oo$ in order for there to be an observable effect on linear cosmology, so all powers in $\phi$ contribute at the same order. Assuming $\Lambda_\star \ll \MPl$ also means that the effective Planck mass seen by gravity is indeed $\MPl$ at leading order, so the $\phi$-dependence of $G_4$ does not strongly affect the background evolution (decoupling discussions of the background dynamics and of the evolution of linear perturbations as before).  
Nevertheless, the special case $\Lambda_\star \sim \MPl$ is qualitatively different and also of interest. Here the interactions \eqref{largeloops2} no longer dominate and a large number of new interactions appear at leading order (in the Einstein frame), so we will leave this case for future work.
IV) We have implicitly assumed that only a few specific dimensionful scales enter the $G_3$ and $G_4$ interactions, finding that loop corrections can in principle become problematically large when the tree level contribution to the $\hat\alpha_i$ is large. 
One may attempt to disentangle tree and loop contributions by fiat, specifically via postulating that the UV completion is weakly coupled at the expense of introducing an additional small dimensionless coupling parameter $g$, which suppresses all loops by construction. For details on this approach see \cite{deRham:2017imi,Adams:2006sv,Burgess:1992gx,Giudice:2007fh}.
Here, however, we have adopted a conservative, minimal approach and refrained from imposing additional  assumptions about the UV, extended symmetries etc.

\bibliographystyle{apsrev4-1}
\bibliography{HRS}

\end{document}